\documentclass[a4paper,11pt]{article}
\usepackage{pos}
\usepackage{braket}
\usepackage{amsmath}
\usepackage{amssymb}
\usepackage{multicol}
\usepackage{slashed}
\usepackage{mathtools}
\usepackage{float}
\usepackage{dsfont}
\newcommand{\Tr}{{\rm Tr}}

\usepackage[utf8]{inputenc}\DeclareUnicodeCharacter{2212}{-}
\usepackage[normalem]{ulem}

\title{Glueballs in $N_f=1$ QCD}

\newcommand{\be}{\begin{equation}}
\newcommand{\ee}{\end{equation}}
\newcommand{\bea}{\begin{eqnarray}}
\newcommand{\eea}{\end{eqnarray}}

\author[a]{Andreas Athenodorou}
\author*[b]{Georg Bergner}
\author[c,d]{Michael~Teper}
\author[e]{Urs Wenger}

\affiliation[a]{Computation-based Science and Technology Research Center, The Cyprus Institute, Cyprus}
\affiliation[b]{University of Jena, Institute for Theoretical Physics, Max-Wien-Platz 1, 07743 Jena, Germany}
\affiliation[c]{Rudolf Peierls Centre for Theoretical Physics, University of Oxford, Parks Road, Oxford OX1 3PU, UK}
\affiliation[d]{All Souls College, University of Oxford, High Street, Oxford OX1 4AL, UK}
\affiliation[e]{Albert Einstein Center for Fundamental Physics, Institute for Theoretical Physics, University of Bern, Sidlerstrasse 5, CH–3012 Bern, Switzerland}

\abstract{\color{black} We present an evaluation of the glueball spectrum for configurations produced with $N_f=1$ dynamical fermions as a function of the $m_{\rm PCAC}$ mass.  We obtained masses of states that fall into the irreducible representations of the octahedral group of rotations in combination with the quantum numbers of charge conjugation $C$ and parity $P$. Due to the low signal to noise ratio, practically, we can only extract masses for the irreducible representations $R^{PC}=$ $A_1^{++}$, $E^{++}$, $T_2^{++}$ as well as $A_1^{-+}$. We make use of the Generalized Eigenvalue Problem (GEVP) with an operator basis consisting only of gluonic operators. Throughout this work we are aiming towards the identification of the effects of light dynamical quarks on the glueball spectrum and how this compares to the statistically more precise spectrum of SU(3) pure gauge theory. We used large gauge ensembles which consist of ${\sim {~\cal O}}(10 {\rm K})$ configurations. Our findings demonstrate that the low-lying spectrum of the scalar, tensor as well as pseudo-scalar glueballs receive negligible contributions from the inclusion of $N_f=1$ dynamical fermions.}

\FullConference{The 40th International Symposium on Lattice Field Theory (Lattice 2023)\\
July 31st - August 4th, 2023\\
Fermi National Accelerator Laboratory\\}

\begin{document}
\maketitle

\section{Introduction}
\label{sec:introduction}
Glueballs are resonance states consisting solely of gluons with a color singlet configuration, a phenomenon anticipated by the confinement principle in Quantum Chromodynamics (QCD). While various possible candidates for glueballs have been detected, a consensus on their precise identification remains elusive, making it one of the unresolved enigmas in the field of hadron spectroscopy. 

Over the past few years, new experimental instruments such as  PANDA~\cite{Parganlija:2013xsa} and BESIII~\cite{Asner:2008nq} have become operational, with additional ones on the horizon. These advancements will yield fresh data and analytical insights into the gluon-rich channels that have been previously explored. This, in turn, will present a challenge to the novel theoretical methodologies and results that have been recently proposed, encompassing both lattice and analytical approaches. Recent reviews on the search for glueballs can be found in the Lattice 2022 plenary presentation by D. Vadacchino~\cite{vadacchino_davide_2022_7338133} as well as in the review by E. Klempt in Ref.~\cite{Klempt:2022ipu}.

{\color{black} Recent results on the glueball spectrum~\cite{Athenodorou:2023ntf} obtained with $N_f=4$ dynamical fermions revealed the existence of an additional state, which manifests as the lightest state in the scalar channel ($A_1^{++}$). This state appears to be associated with the decay of a glueball to two or four pions. It would, thus, be useful to investigate the effect of light quarks in a theory were such decays are suppressed but dynamical fermions can still affect the nature of the spectrum. Such a case is $N_f~=~1$ QCD where pions do not exist.}

In this study, we aim to delve into the impact of a single light fermion on the glueball spectrum. To achieve this goal, we utilize configurations generated with a single light Clover quark ($N_f = 1$) across a range of bare masses. We extract the glueball spectrum and then compare it with that obtained from pure gauge SU(3) configurations produced with the Symanzik tree-level improved action at two values of the gradient flow. For massive dynamical quarks we expect, from decoupling arguments, that the glueball spectrum becomes similar to the spectrum of the pure gauge theory~\cite{Athenodorou:2020ani, Athenodorou:2021qvs}. The important question that arises here is what happens if one includes light dynamical fermions.

Overall, our main finding of the investigation of $N_f=1$ QCD is that the spectrum, 
at the given statistical accuracy of ${\cal O} (10 {\rm K})$ configurations, appears to be consistent with the pure gauge theory and independent of the fermion mass. 

{\color{black} This manuscript is structured as follows. We begin in Section~\ref{sec:simulation_details} by presenting the lattice setup used to generate configurations with $N_f=1$, along with those using the pure gauge action. Moving on to Section~\ref{sec:glueball_masses}, we provide a brief explanation of how the glueball spectrum in Lattice QCD can be extracted using the Generalized Eigenvalue Problem (GEVP) method. Next, in Section~\ref{sec:topological_charge_and_scale_setting}, we describe the process of calculating the topological charge, which serves as a measure of the system's ergodicity. We also detail the evaluation of the energy scale $t_0$ through the smoothing scheme of the gradient flow. Subsequently, we focus on presenting the results, specifically discussing the scalar channel $R^{PC}=A_1^{++}$, the tensor  $R^{PC}=E^{++}$ and $T_2^{++}$ channels, and the pseudoscalar glueball obtained in the $R^{PC}=A_1^{-+}$ channel. Finally, we conclude the proceedings in Section~\ref{sec:conclusions}.}

\section{Simulation Details}
\label{sec:simulation_details}
The lattice configurations have been generated as part of an extension of a larger project focused on one flavour QCD started by the DESY-Münster collaboration \cite{Farchioni:2006waf,Farchioni:2007dw,Farchioni:2008na}. The first ensembles have been generated with a tree-level Symanzik-improved gauge action and one level of stout smearing in the standard Wilson fermion action. This has been later extended to a tree-level clover-improved fermion action. While the first configurations have been generated with the polynomial hybrid Monte-Carlo algorithm, later on the rational hybrid Monte-Carlo algorithm has been used from a newly developed code package. The meson masses including the $\eta_S$ and $\sigma_S$ have been determined at earlier stages of the project. The mixing of meson and glueball operators has only been considered in a very preliminary study. Here we report on a considerable update of the glueball sector using the improved fermion action. For our analysis, we have selected ensembles at $\beta=4.2$ and $\beta=4.4$. To facilitate a comparison, we also conducted simulations of pure SU(3) gauge theory with  tree-level Symanzik-improved gauge action using the hybrid Monte-Carlo algorithm. The simulated values of $\beta$ are $\beta=4.51$ and $\beta=4.75$, corresponding respectively to the $\beta=4.2$ and $\beta=4.4$ values employed in the $N_f=1$ simulation with clover improvement.


\section{Calculation of glueball masses}
\label{sec:glueball_masses}
Glueball masses can be established through the use of the standard decomposition technique applied to a Euclidean correlator involving an operator denoted as $\phi(t)$. This decomposition process relies on representing these physical states within the context of the system's Hamiltonian, denoted as $H$, and the associated energy eigenstates:
\begin{eqnarray}
\langle \phi^\dagger(t=an_t)\phi(0) \rangle
 = 
\langle \phi^\dagger e^{-Han_t} \phi \rangle
= \sum_i |c_i|^2 e^{-aE_in_t} 
\stackrel{t\to \infty}{=} 
|c_0|^2 e^{-aE_0n_t}\,,
\label{extract_mass}
\end{eqnarray}
where $E_0$ represents the ground state energy. The above summation is limited to states that exhibit non-zero overlaps and satisfy the condition $c_i = \langle {\rm vac} | \phi^\dagger | i \rangle \neq 0$. The quantum properties of the operator $\phi$ shall align with those of the particular state being examined. The identification of the ground state hinges on two critical elements: the strength of its correlation with this state and the speed at which we witness exponential decay as outlined in Eq.~(\ref{extract_mass}). Enhancing this correlation entails creating operators that adeptly encapsulate the fundamental characteristics of the state. To extract excited states we employ the GEVP technique~\cite{Luscher:1984is,Luscher:1990ck,Berg:1982kp} applied to a set of operators $\phi_i$ formed from various lattice loops at different blocking levels~\cite{Lucini:2004my,Teper:1987wt}. This involves using correlation matrices, denoted as $C_{ij} = \langle \phi_i^{\dagger} (t) \phi_j (0) \rangle$, where $i,j=1,...,N_{\rm op}$, in conjunction with GEVP. Here, $N_{\rm op}$ represents the number of operators used.

To construct an operator that projects onto a glueball state, we create an ordered product of SU(3) link matrices along a loop that can be continuously contracted and then calculate its trace. The real (imaginary) part of this trace corresponds to positive (negative) charge conjugation $C=+$($-$). In order to ensure that the operator possesses zero momentum, we sum over all spatial translations of the loop. Additionally, we account for all conceivable rotations of the loop and combine them in ways that adhere to the irreducible representations ($R$) of the rotational symmetry group. To create operators with both parities ($P=\pm$), we construct the parity inverse for each loop and then take appropriate linear combinations. In Fig.~\ref{fig:glueball_operators}, we provide a selection of the paths used in constructing our basis.
\begin{center}
\begin{figure}[h]
{\includegraphics [width=75mm] {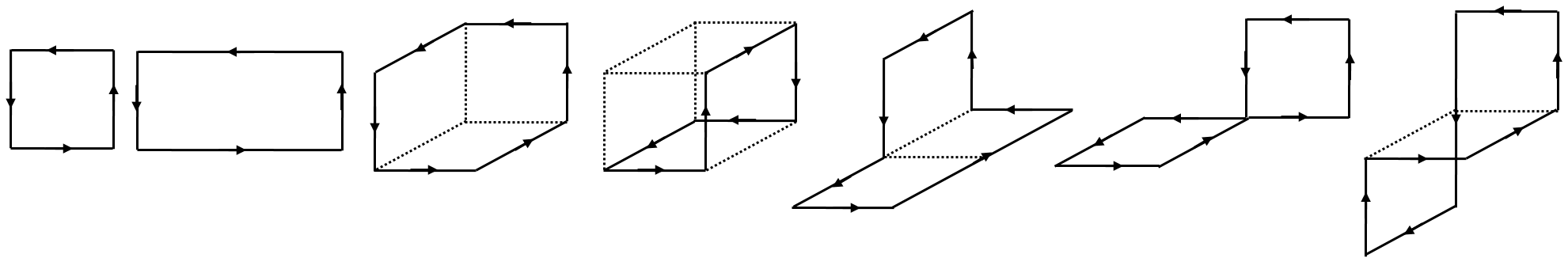}\includegraphics [width=75mm] {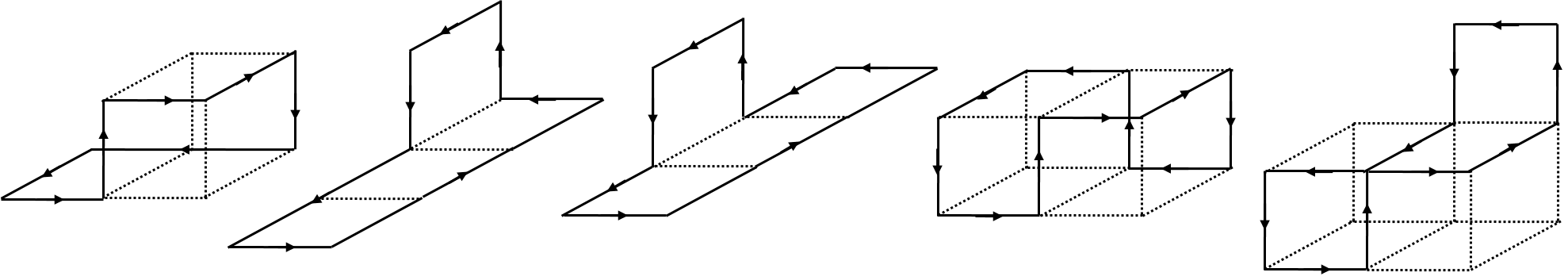}}
\caption{All the different closed loops used for the construction of the glueball operators.}
    \label{fig:glueball_operators}
\end{figure}
\vspace{-1cm}
\end{center}

The irreducible representations $R$ of the discrete subgroup of cubic rotations within the full rotation group are denoted as $A_1, A_2, E, T_1, T_2$. The $A_1$ representation is a singlet and possesses full cubic rotational symmetry, thereby encompassing the $J=0$ state in the continuum limit. Similarly, the $A_2$ representation is also a singlet. The $E$ representation forms a doublet, while both $T_1$ and $T_2$ representations are triplets. In the lattice setting, the three states corresponding to the triplet of $T_2$ are degenerate. To address this, we average their values and treat them as a single state when estimating glueball masses. The same procedure is applied to the $E$ doublets, where their mass estimates are averaged.

The representations of rotational symmetry described above are based on our cubic lattice formulation. As we approach the continuum limit, these states will converge to continuum glueball states that belong to representations of the continuous rotational symmetry. Consequently, they will fall into degenerate multiplets consisting of $2J + 1$ states, where $J$ represents the spin of the states. When determining the continuum limit of the low-lying glueball spectrum, it is more valuable to assign states to a specific spin $J$, rather than to representations of the cubic subgroup, which provide a less precise 'resolution' as they map all spins $J = 1, 2, 3, \dots, \infty$ to just 5 cubic representations. For low values of $J$ ($J=0,1,2$), the distribution of the $2J + 1$ states can be characterized as $A_1 \to J=0$, $T_1 \to J=1$,  and $E, T_2 \to J=2$. 

\section{Topological charge and scale setting}
\label{sec:topological_charge_and_scale_setting}
\begin{figure}[h]
    \vspace{-0.25cm}
    \includegraphics[width=\textwidth]{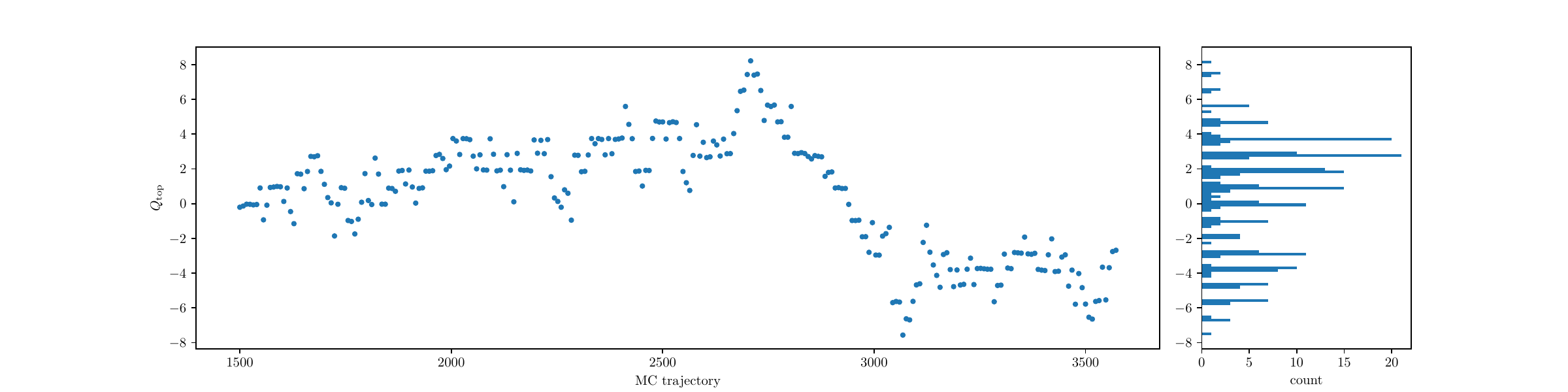}
    \includegraphics[width=\textwidth]{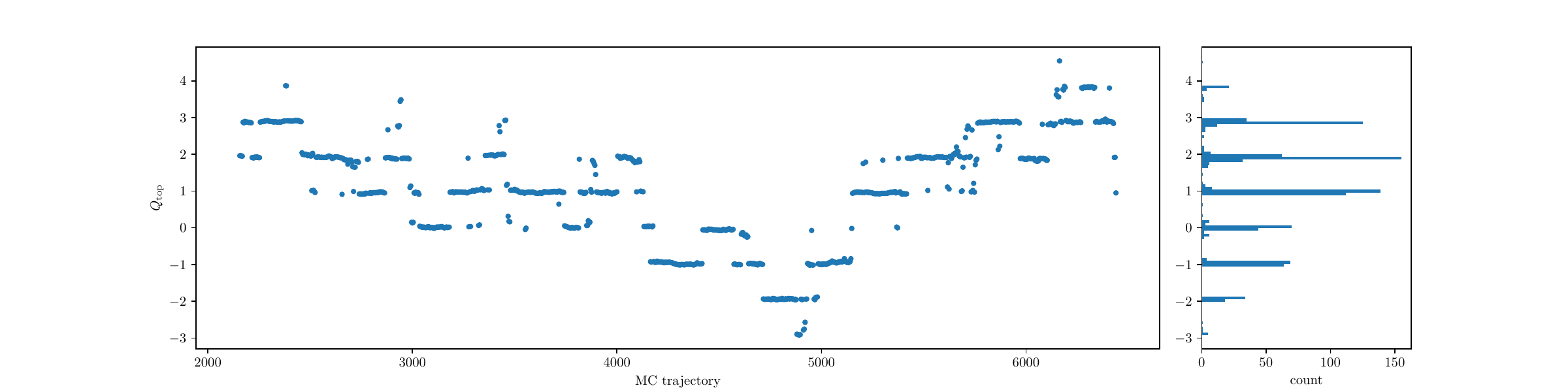}
    \caption{Topological charge fluctuations as a function of the Monte-Carlo time for an ensemble at $\beta=4.2$ at flow time $t/a^2=8.0$ (above) and $\beta=4.4$ at flow time $t/a^2=10.0$ (below).}
    \label{fig:topological_charge}
\end{figure}

{
In the continuum limit, the topological charge is established as the integral across the entire four-dimensional Euclidean space-time volume of the topological charge density $Q = \frac{1}{32\pi^2} \int d^4 x \: \epsilon_{\mu\nu\rho\sigma} \Tr\left[F_{\mu\nu}(x)F_{\rho\sigma}(x)\right] \,.$ 
We employed a lattice version of $Q$ known as the symmetric or 'clover' definition, which was initially introduced in  Ref.~\cite{DiVecchia:1981aev}.
We use the gradient flow~\cite{Luscher:2010iy} in order to smooth out the UV fluctuations of the gauge field defining the topological charge. The smoothing action utilized in the flow equation is the standard Wilson action. 


In Fig.~\ref{fig:topological_charge} we present the history of the topological charge as well as its distribution for two $N_f=1$ ensembles, namely one at $\beta=4.2$ and one at $\beta=4.4$. Clearly, the plots do not indicate severe topological freezing, suggesting that the Markov-Chain is ergodic. 

The gradient flow technique additionally allows for the establishment of a well-defined physical scale parameter denoted as $t_0$, which can be determined with a high degree of precision. This concept of $t_0$ was originally introduced in references \cite{Luscher:2009eq, Borsanyi:2012zs}. The definition of $t_0$ follows a specific prescription as outlined below. First, we set $F(t) = t^2 \langle E(t) \rangle \, \ {\rm with} \ E(t) = \frac{1}{4} B^2_{\mu \nu} (t)\,,$ where $B_{\mu \nu}$ is field strength obtained by flowing $F_{\mu \nu}$ along the flow time direction. We define the scale $t_0(c)$ as the value of $t$ for which $F(t) |_{t=t_0(c)} = c\,$ where $c$ should be chosen so that the relevant condition $a \ll \sqrt{8 t_0} \ll L$ is satisfied. Small values of $c$ lead to large lattice artefacts while large $c$ usually lead to larger autocorrelations \cite{Bergner:2014ska}. In our case we choose the value $c=0.3$ which is the value commonly used in lattice QCD calculations.}

\section{Results}
\label{sec:results}
\begin{figure}[h]
     \vspace{-0.35cm}    
    \centering
    \scalebox{0.95}{\rotatebox{0}{\hspace{-0.250cm}\includegraphics[width=165mm]{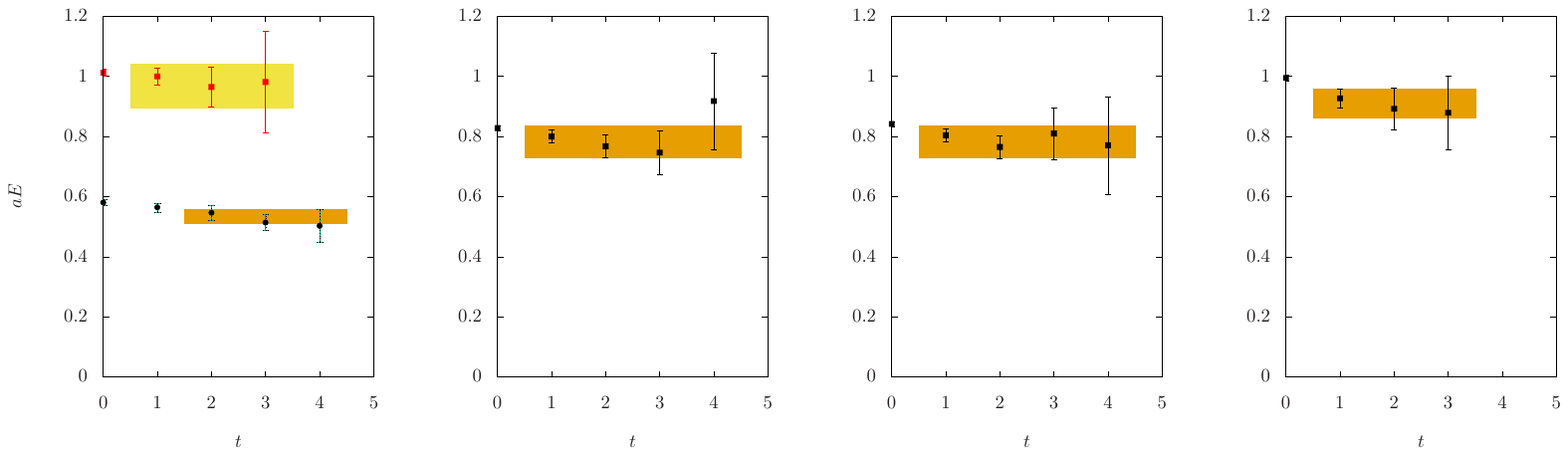}\put(-380,330){$A_1^{++}$}\put(-270,330){$E^{++}$}\put(-155,330){$T_2^{++}$}\put(-45,330){$A_1^{-+}$}}}
     \vspace{-8.00cm}
    \caption{Here we provide examples of well-defined mass plateaus, along with their most accurate estimates for the four irreducible representations. In the plots above, moving from left to right, we showcase effective mass plots for: the ground and first excited states of $A_1^{++}$, one of the doublets of the $E^{++}$ ground state, one of the triplets in the $T_2^{++}$ ground state, the $A_1^{-+}$ ground state.}
    \label{fig:plots_effective_masses_glueballs}
\end{figure}

We have successfully acquired the low-energy spectra associated with the irreducible representations $A_1^{++}$, $E^{++}$, and $T_2^{++}$, as well as $A_1^{-+}$, which correspond to the scalar, tensor, and pseudoscalar channels respectively. An intriguing observation stemming from our calculations is the early establishment of effective mass plateaus, illustrated in Figure~\ref{fig:plots_effective_masses_glueballs}, in stark contrast to what has been observed in the case of $N_f=4$, where the plateaus set in later during the temporal evolution. These results are marked by high overlaps ranging from 80\% to 100\%. Notably, this phenomenon closely resembles the rapid convergence of mass plateaus observed in the context of SU(3) pure gauge theory. It could potentially indicate a significant reduction in the number of states appearing in the aforementioned Hilbert space of the $N_f=1$ QCD vacuum compared to that of $N_f=4$ QCD. Consequently, the quality of the mass plateau appears to be akin to that of a pure gauge theory.

\begin{figure}[h]
\vspace{-0.25cm} 
    \centering
\scalebox{0.95}{\rotatebox{0}{\hspace{-0.25cm}\includegraphics[width=165mm]{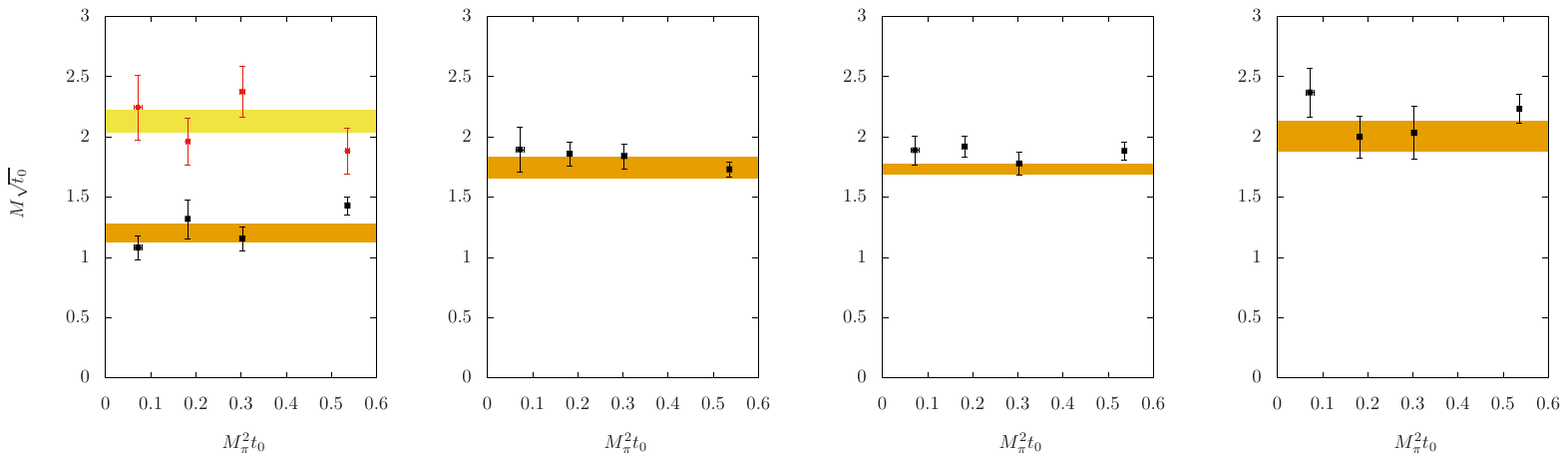}\put(-380,330){$A_1^{++}$}\put(-270,330){$E^{++}$}\put(-155,330){$T_2^{++}$}\put(-45,330){$A_1^{-+}$}\put(-390,275){\tiny g.s}\put(-400,302){\color{red} \tiny $1^{\rm st}$ ex.s}\put(-270,292){\tiny g.s}\put(-153,292){\tiny g.s}\put(-40,298){\tiny g.s}}}
\vspace{-7.75cm}
    \caption{Glueball masses for $N_f=1$ and $\beta=4.4$ improved case vs.~the PQChPT pion mass. The bands represent the mass estimates for SU(3) pure gauge theory for $\beta=4.75$.}
    \label{fig:plots_Nf1_beta_4.4_improved}
\end{figure}

In Figure~\ref{fig:plots_Nf1_beta_4.4_improved} we provide results of the glueball masses in units of $1/\sqrt{t_0}$ as a function of the PQChPT pion mass for the $(i)$ ground and first excited states for $A_1^{++}$, $(ii)$ the ground state for $E^{++}$, $(iii)$ the ground state for $T_2^{++}$, and $(iv)$ the ground state for $A_1^{-+}$. The parameters of the ensembles used to produce the aforementioned plot are $\beta=4.4$, and $\kappa=0.1280, 0.1287, 0.1290$ and $0.1293$. The bands represent the mass estimates for SU(3) pure gauge theory for $\beta=4.75$ which corresponds to $t_0/a^2 \sim 7.07$. The above value of $t_0/a^2$ matches the corresponding values for the $N_f=1$ ensembles at $\beta=4.4$. The level of agreement between the results for $N_f=1$ and the pure gauge theory of SU(3) is astonishing, demonstrating that the effects resulting from the inclusion of the dynamical quark into the vacuum are negligible at the given level of accuracy. Hence, the glueball masses, are independent of the quark mass. 

\begin{figure}[h]
    \centering
\scalebox{0.95}{\rotatebox{0}{\hspace{-0.25cm}\includegraphics[width=165mm]{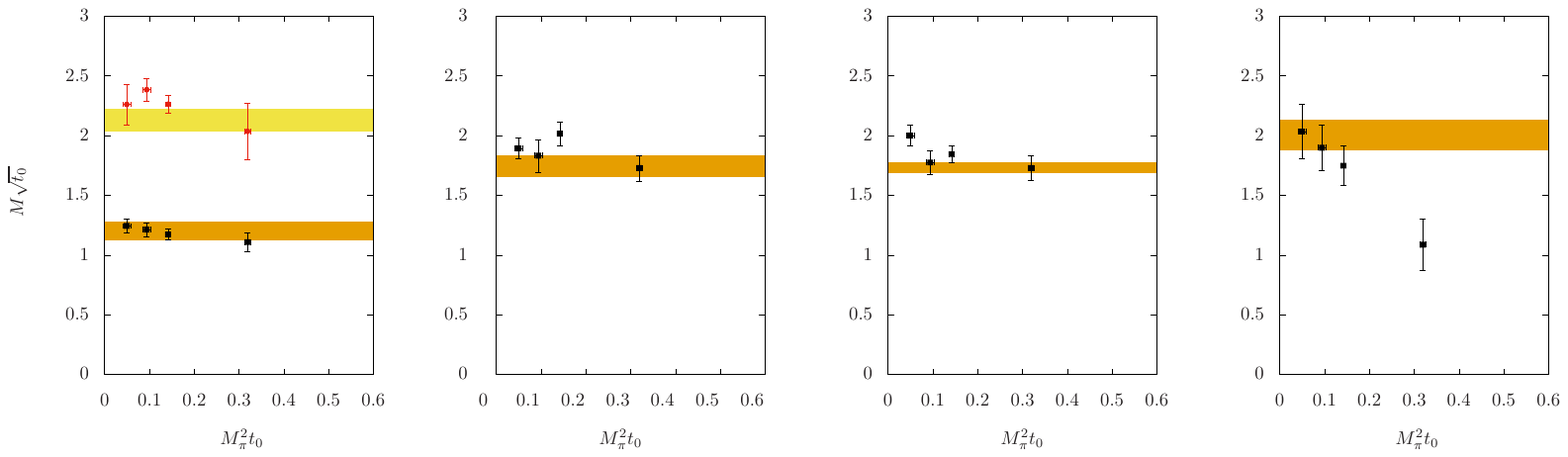}\put(-380,330){$A_1^{++}$}\put(-270,330){$E^{++}$}\put(-155,330){$T_2^{++}$}\put(-45,330){$A_1^{-+}$}\put(-370,275){\tiny g.s}\put(-380,302){\color{red} \tiny $1^{\rm st}$ ex.s}\put(-270,292){\tiny g.s}\put(-153,292){\tiny g.s}\put(-40,298){\tiny g.s}}}
\vspace{-7.75cm}
    \caption{Glueball masses for $N_f=1$ and $\beta=4.4$ unimproved case vs.~the PQChPT pion mass. The bands represent the mass estimates for SU(3) pure gauge theory at $\beta=4.75$.}
    \label{fig:plots_Nf1_beta_4.4_unimproved}
\end{figure}

An investigation of the same theory at $\beta=4.4,$ without ${\cal O}(a)$ fermionic improvement, reveals a similar pattern, as shown in Figure~\ref{fig:plots_Nf1_beta_4.4_unimproved}. However, it is important to note that while the results for the irreducible representations $A_1^{++}$, $E^{++}$, and $T_2^{++}$ are independent of the quark mass, the glueball mass for the ground state of the pseudoscalar channel $A_1^{-+}$ appears to decrease with increasing quark mass. This effect disappears in the improved case, leading us to interpret this behavior as a consequence of lattice artifacts. While $t_0/a^2 \approx 5.2$ in the unimproved theory, the bands in  Figure~\ref{fig:plots_Nf1_beta_4.4_unimproved} once again denote the mass estimates for SU(3) pure gauge theory at $\beta=4.75.$ This assumes that lattice artifacts on $M \sqrt{t_0}$ for the pure gauge theory exhibit negligible differences between $t_0/a^2 \approx 5.2$ and $7.07$.
 \section{Conclusions}
\label{sec:conclusions}
The spectrum of $N_f=1$ QCD, at the given statistical accuracy of ${\cal O} (10 {\rm K})$ configurations, appears to be consistent with the one of the pure gauge theory and independent of the fermion mass with no any other states showing up at low energies. This has been confirmed for two values of $\beta$ as well as for ${\cal O}(a)$ improved vs.~unimproved fermionic descretizations. This suggests that the effects of one dynamical fermion on the glueball spectrum are insignificant. In the future, we will also consider mesonic operators to investigate possible mixings between glueballs and mesons.

\section{Acknowledgements}
Calculations were performed on the Cyclone HPC system at The Cyprus Institute as well as on  UBELIX, the HPC cluster at the University of Bern. AA received financial support by the EuroCC2 project funded by the Deputy Ministry of Research, Innovation and Digital Policy and the Cyprus
Research and Innovation Foundation and the European High-Performance Computing Joint Undertaking (JU) under grant agreement No 101101903. MT acknowledges support by the Simons Collaboration on
Confinement and QCD Strings.
\bibliographystyle{JHEP}
\bibliography{biblio_NEW}
\end{document}